\documentclass{jfm}

\usepackage{graphicx}
\usepackage{natbib}
\usepackage{epsf}

\ifCUPmtlplainloaded \else
  \checkfont{eurm10}
  \iffontfound
    \IfFileExists{upmath.sty}
      {\typeout{^^JFound AMS Euler Roman fonts on the system,
                   using the 'upmath' package.^^J}%
       \usepackage{upmath}}
      {\typeout{^^JFound AMS Euler Roman fonts on the system, but you
                   dont seem to have the}%
       \typeout{'upmath' package installed. JFM.cls can take advantage
                 of these fonts,^^Jif you use 'upmath' package.^^J}%
      }
  \else
  \fi
\fi

\ifCUPmtlplainloaded \else
  \checkfont{msam10}
  \iffontfound
    \IfFileExists{amssymb.sty}
      {\typeout{^^JFound AMS Symbol fonts on the system, using the
                'amssymb' package.^^J}%
       \usepackage{amssymb}%
         
       \let\ge=\geqslant  \let\geq=\geqslant
      }{}
  \fi
\fi

\ifCUPmtlplainloaded \else
  \IfFileExists{amsbsy.sty}
    {\typeout{^^JFound the 'amsbsy' package on the system, using it.^^J}%
     \usepackage{amsbsy}}
    {\providecommand\boldsymbol[1]{\mbox{\boldmath $##1$}}}
\fi

\newsavebox{\astrutbox}
\sbox{\astrutbox}{\rule[-5pt]{0pt}{20pt}}

\newcommand\be{{\ensuremath e}}

\newcommand\bu{{\ensuremath u}}
\newcommand\bv{{\ensuremath v}}

\newcommand\bOmega{{\ensuremath\Omega}}
\newcommand\bnabla{{\ensuremath\nabla}}
\newcommand\rmd{\mathrm{d}}
\newcommand\rme{\mathrm{e}}
\newcommand\rmi{\mathrm{i}}
\newcommand\f{\frac}
\newcommand\p{\partial}

\newcommand\rma{\mathrm{a}}
\newcommand\rmb{\mathrm{b}}
\newcommand\rmg{\mathrm{g}}
\newcommand\re{\mathrm{Re}}

\title[Inertial Wave Attractors]{Direct numerical simulations of an inertial wave attractor in linear and nonlinear regimes}

\author[L. Jouve and G. I. Ogilvie]%
{Laur\`ene Jouve$^1$%
  \thanks{Email address for correspondence: ljouve@irap.omp.eu} and Gordon I. Ogilvie$^2$}

\affiliation{$^1$UPS-OMP, Institut de Recherche en Astrophysique et Plan\'etologie, Universit\'e de Toulouse CNRS, 14 Avenue Edouard Belin, 31400 Toulouse, France\\[\affilskip]
$^2$Department of Applied Mathematics and Theoretical Physics, University of Cambridge, Centre for Mathematical Sciences, Wilberforce Road, Cambridge CB3 0WA, UK}

\pubyear{}
\volume{}
\pagerange{}
\date{?; revised ?; accepted ?. - To be entered by editorial office}
\begin{document}

\maketitle

\begin{abstract}
  In a uniformly rotating fluid, inertial waves propagate along rays
  that are inclined to the rotation axis by an angle that depends on
  the wave frequency. In closed domains, multiple reflections from the
  boundaries may cause inertial waves to focus on to particular
  structures known as wave attractors. These attractors are likely to
  appear in fluid containers with at least one boundary that is
  neither parallel nor normal to the rotation axis. A closely related
  process also applies to internal gravity waves in a stably
  stratified fluid. Such structures have previously been studied from
  a theoretical point of view, in laboratory experiments, in linear
  numerical calculations and in some recent numerical simulations. In
  the present paper, two-dimensional direct numerical simulations of
  an inertial wave attractor are presented. By varying the amplitude
  at which the system is forced periodically, we are able to describe
  the transition between the linear and nonlinear regimes as well as
  the characteristic properties of the two situations. In the linear
  regime, we first recover the results of the linear calculations and
  asymptotic theory of \cite{Ogilvie05} who considered a prototypical
  problem involving the focusing of linear internal waves into a
  narrow beam centred on a wave attractor in a steady state. The
  velocity profile of the beam and its scalings with the Ekman number,
  as well as the asymptotic value of the dissipation rate, are found
  to be in agreement with the linear theory. We also find that, as the
  beam builds up around the wave attractor, the power input by the
  applied force reaches its limiting value more rapidly than the
  dissipation rate, which saturates only when the beam has reached its
  final thickness. In the nonlinear regime, the beam is strongly
  affected and becomes unstable to a subharmonic instability. This
  instability transfers energy to secondary waves possessing shorter
  wavelengths and lower frequencies. The onset of the instability of
  a narrow inertial wave beam is investigated by means of a separate
  linear analysis and the results, such as the onset of the
  instability, are found to be consistent with the global simulations
  of the wave attractor. The excitation of such secondary waves
  described theoretically in this work has also been seen in recent
  laboratory experiments on internal gravity waves.
\end{abstract}

\section{Introduction}

Internal waves supported by buoyancy and Coriolis forces play an
important role in the Earth's ocean and atmosphere, and have been
widely studied through theory and experiment.  The simplest situations
are those of pure internal gravity waves in a non-rotating, stably
stratified fluid, and pure inertial waves in a rotating, unstratified
fluid.  There is a close analogy between these two cases.

Internal waves have properties quite different from acoustic or
electromagnetic waves, being highly anisotropic and dispersive.  In a
frame of reference in which the mean flow vanishes, the wave frequency
depends only on the direction, and not on the magnitude, of the
wavevector.  Special directions are defined by gravity or rotation.
Energy propagates at the group velocity, which is perpendicular to the
wavevector and inversely proportional to its magnitude.  
Internal waves are
highly dispersive and do not steepen like acoustic waves, but instead break and dissipate
when they exceed a critical amplitude and become locally unstable. Their frequencies are bounded and form a
dense or continuous spectrum.  In a closed domain, internal waves
generally do not form normal modes with regular eigenfunctions,
although the simplest cases are exceptions to this rule.  Waves of a
given frequency generally undergo focusing or defocusing when they
reflect from boundaries; after multiple reflections in a generic
container, rays are typically focused towards limit cycles known as wave
attractors \citep{Maas95}.

In the interiors of stars and giant planets, internal waves occupy the low-frequency part of
the spectrum of oscillations, which can be excited by tidal forcing
when the body has an orbiting companion \citep[e.g][]{Ogilvie04}.  The
question arises of how a rotating or stably stratified fluid responds
to an oscillatory body force with a frequency that lies within the
dense or continuous spectrum.  For tidal applications, what is of
particular interest is the rate at which energy is transferred to the
fluid, ultimately to be dissipated.  The relative amplitudes of
astrophysical tides can vary enormously, depending on the relative
mass and distance of the tide-raising body.  It is therefore
appropriate to investigate both linear and nonlinear regimes for
internal tides.

\cite{Ogilvie05} considered a simple problem involving a two-dimensional
fluid domain that supports internal waves, and calculated the
steady-state linear response to a periodic body force with a frequency
such that a wave attractor occurs.  He obtained an analytical
asymptotic solution in the limit that the dissipative effects of the
fluid are small. The response is localized in a narrow beam centred on the attractor, 
and, for a body force of fixed amplitude, the dissipation rate (which is equal to the power input) 
is asymptotically independent of both the magnitude and the details of the dissipative process. 
This situation occurs because the attractor acts as a `black
hole' that absorbs a certain energy flux from the waves being focused
into it. This result contrasts strongly with a situation in which 
the internal waves form classical normal modes, as in the case of inertial 
waves in a rectangular container whose walls are parallel and perpendicular to the rotation axis. 
In such a situation, the dissipation rate is simply proportional to the viscosity, and therefore 
astrophysically unimportant, unless the forcing frequency matches the natural frequency of 
a normal mode and a resonance occurs.

Internal wave attractors have been seen in experiments
\citep{Maas97,Maas01,Manders03, Lam08, Hazewinkel08, Hazewinkel11a,
  Hazewinkel11b,Scolan13}, linear calculations \citep{Rieutord97,
  Dintrans99, Rieutord01, Ogilvie05, Maas07, Ogilvie09, Rieutord10,
  Echeverri11, Bajars13, Baruteau13} and numerical simulations
\citep{Grisouard08}.  The wave attractor is a linear phenomenon and
one may naturally ask how it is modified by nonlinearity.  Previous
experiments \citep{Maas01} have shown some evidence for cumulative
nonlinear effects that build up over a large number of wave periods:
the fluid becomes increasingly mixed, and momentum deposition leads to
the generation of a mean flow.  Until recently, no evidence had been
found for stronger nonlinear phenomena such as wave breaking,
instability or turbulence.  However, \citet{Scolan13} have recently
reported experiments on internal gravity waves in which waves focused
towards an attractor become unstable and divert energy to smaller
scales.  They describe this process as a parametric subharmonic
instability similar to that known for plane internal waves.

The purpose of this work is to study a simple system of periodically
forced inertial waves in a closed domain by means of fully nonlinear
direct numerical simulations.  We aim first to recover the results of
\cite{Ogilvie05} for the steady-state response of a wave attractor in
a linear regime, and to understand how this steady state is achieved.
Moving into the nonlinear regime, we aim to locate the onset of
instability and to see how the attractor operates, if at all, in the
presence of instability.  We will compare our results with the
experimental findings of \citet{Scolan13}.  Of particular interest is
the energy budget of the system and the question of how the dissipation rate 
(and thence the tidal torque, in astrophysical applications) is affected by nonlinearity.

\section{Model and governing equations}

\subsection{Description of the model}
\label{sect_intro}

\begin{figure}
\centering
\includegraphics[width=6cm]{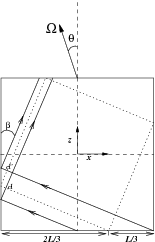}
\caption{Sketch of the fluid domain and wave attractor. The dotted
  line represents the theoretical spatial structure of the inertial
  wave attractor in this situation. The focusing of one beam of width
  $d$ is shown for one reflection at the left boundary. The resulting
  beam has width $d^{\prime}=d/2$.}
\label{fig1}
\end{figure}

In this section, we present the model system that we use to study the
forcing of inertial waves in a closed two-dimensional (2D) domain.  A
similar model was used by \cite{Ogilvie05} for analytical and numerical
calculations of forced linear internal waves.  Perhaps the simplest
domain exhibiting a wave attractor is a square that is tilted with
respect to the rotation axis.  We therefore adopt Cartesian
coordinates $(x,y,z)$ in a rotating frame of reference and solve the
incompressible Navier--Stokes equation in a square domain
$-L/2<\{x,z\}<L/2$ with the angular velocity vector
$\boldsymbol{\Omega}$ in the $xz$~plane and inclined by an angle
$\theta$ with respect to the $z$~axis (Fig.~\ref{fig1}).  The problem is
invariant and unbounded in the $y$~direction and the fluid velocity
$\boldsymbol{v}$ has three components.

For wave frequencies smaller than twice the rotation frequency
($|\omega|<2\Omega$), the spatial structure of inviscid linear waves
in the $xz$~plane is described by a hyperbolic equation (the
Poincar\'e equation), whose characteristic rays are straight lines
inclined at angles $\pm\arcsin(\omega/2\Omega)$ with respect to the
rotation axis.  Indeed, the dispersion relation relates the frequency
and the direction of propagation (which is perpendicular to the
wavevector $\boldsymbol{k}$) through $\omega=\pm2\boldsymbol{\hat
  k}\cdot\boldsymbol{\Omega}$, where $\boldsymbol{\hat
  k}=\boldsymbol{k}/|\boldsymbol{k}|$.

The rotation axis in our model is inclined with respect to the
boundaries so that, allowing for multiple reflections from the
walls, the rays tend to focus on to localized singular structures
known as wave attractors \citep{Maas95}.  (In the case of aligned
rotation, i.e.\ $\theta=0$, the rays instead propagate ergodically
through the domain, except for special frequencies for which they
propagate periodically.)
A body force is applied to the system at a particular temporal
frequency (less than $2\Omega$) to mimic the effect of tidal forcing.
In order to excite waves in an incompressible fluid in a closed
domain, a non-potential force must be applied, and the origin of this
vortical effective force in a tidal problem with free boundaries is
explained in \cite{Ogilvie05}.  Contrary to the turbulent wave
excitation in a convective environment, for example, tidal forcing is
strictly periodic (or nearly so) and thus can be easily imposed in our
equation of motion.

The governing equations 
for a homogeneous viscous incompressible fluid in a rotating frame are then
\begin{equation}
  \frac{\partial \boldsymbol{v}}{\partial t} + (\boldsymbol{v} \cdot \boldsymbol{\nabla}) \boldsymbol{v} + 2\boldsymbol{\Omega} \times \boldsymbol{v} = -\boldsymbol{\nabla} P + \nu \, \nabla^2 \boldsymbol{v} + \boldsymbol{a},
\label{eq_NS}
\end{equation}
\begin{equation}
  \nabla \cdot \boldsymbol{v} = 0,
\label{eq_div}
\end{equation}
where $P$ is a modified pressure that incorporates the gravitational
and centrifugal potentials, and $\boldsymbol{a}$ is the body force per
unit mass. As explained above, only the rotational part of
$\boldsymbol{a}$ drives the fluid motion in a closed container. We
adopt a simple expression for $\boldsymbol{a}$, linear in the
Cartesian coordinates, such that
\begin{equation}
  \boldsymbol{\nabla}\times\boldsymbol{a} = F_0  \cos(\omega_0 t) \,\boldsymbol{e}_y
\end{equation}
is uniform, as in the numerical example studied by \cite{Ogilvie05}.

We use $1/2\Omega$ as the unit of time and $L$, the size of the domain
in which the fluid is contained, as the unit of length. 
The rotation vector is inclined as shown in Fig.~\ref{fig1}, leading to the following expression for {\bf$\Omega$}:
\begin{equation}
2\boldsymbol{\Omega}=(\cos\theta\,\boldsymbol{e}_z- \sin\theta\,\boldsymbol{e}_x )
\end{equation}

For all the simulations presented here, we choose $\omega_0/2
\Omega=\sin(\pi/4)=\sqrt{2}/2$ so the angle of wave propagation with
respect to the rotation axis is $\pi/4$.  We also choose the
inclination angle $\theta$ such that $\tan\theta=1/3$, which results
in a theoretically square wave attractor inside the square domain,
reflecting from each wall one-third of the way along its length $L$
(Fig.~\ref{fig1}).  

The remaining dimensionless parameters to be chosen are the Ekman
number $Ek=\nu/2\Omega L^2$ and the dimensionless forcing amplitude
$F_0/(2\Omega)^2$.  We work with dimensionless variables, choosing units
such that $L=2\Omega=1$, so that $Ek=\nu$.  

A sketch of the geometry is shown in Fig.~\ref{fig1}. The square
domain is represented together with the inclined rotation vector. An
illustrative example of one reflection on the left boundary of an
inviscid inertial wave beam is shown, where the focusing effect is
quite clear.  It is easy to show that in this situation, the focusing
factor for one reflection is 2, i.e. $d^{\prime}=d/2$ with the
notations chosen here.

\subsection{Numerical implementation}

To solve our problem numerically, we use SNOOPY, a 3D spectral code
solving the equations of incompressible or Boussinesq
(magneto)hydrodynamics using a Fourier representation \citep{Lesur05,
  Lesur07, Lesur10}. We here compute the evolution of the velocity
field in 2D (in the $xz$ plane) by putting only one point in the $y$
direction.  The spatial computational domain is
$-L_\mathrm{c}/2<\{x,z\}<L_\mathrm{c}/2$ and the typical resolution we
use is $256\times256$. To impose rigid, no-slip boundary conditions at
the four edges of the fluid domain $-L/2<\{x,z\}<L/2$, we apply a mask
in physical space on the velocity field so that all three components
of the velocity are forced to vanish in the part of the domain outside
the fluid, which is such that $L/L_\mathrm{c}=0.8$. By applying this
mask in the physical space after each sub-timestep of the third-order
Runge--Kutta method, we inevitably have a loss of energy in the
neighbourhood of the boundaries. Both increasing the spatial resolution
(to resolve the boundary layers) and decreasing the time step (to
decrease the error linked to the mask) lead to a convergence towards
zero of the loss of energy. In the cases shown here, a resolution of
$256$ collocation points per dimension was found to be enough to
ensure energy conservation to an accuracy of a few percent.

The initial condition we use is zero-velocity but, since a body force
is applied to the system, a velocity field rapidly sets in, consisting
of inertial waves whose temporal evolution we wish to follow in
various situations. We start with the linear cases to which analytical
results can be compared and then increase the amplitude of the force
to trigger nonlinearities in the system.

\begin{figure*}
\centering
\includegraphics[width=12cm]{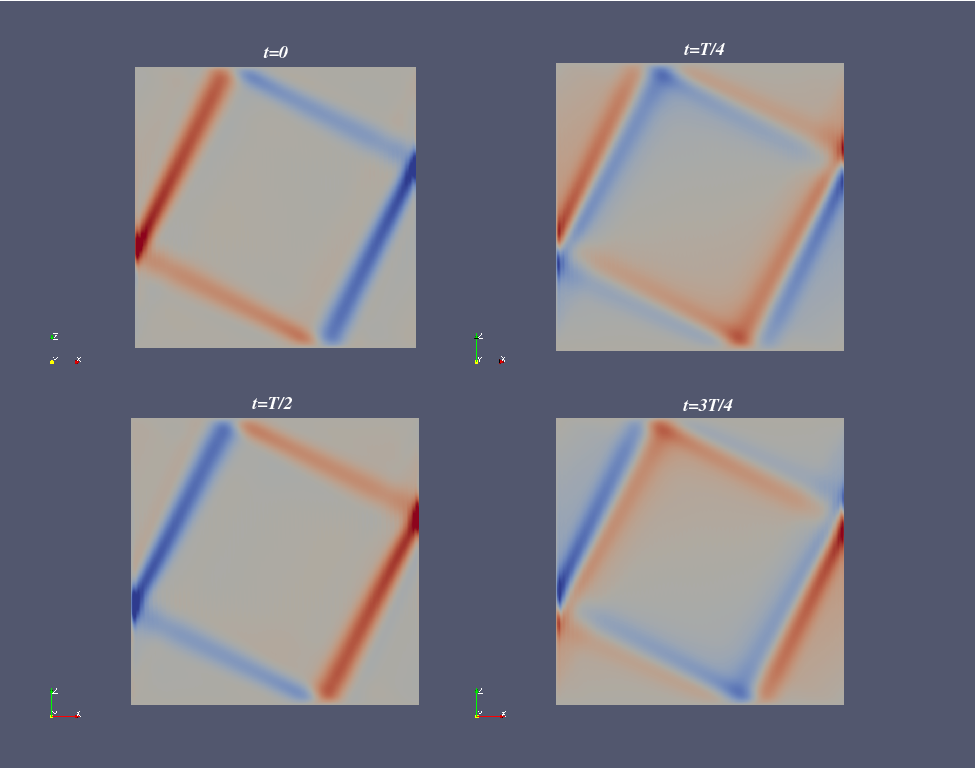}
\caption{Snapshots of $v_z$ inside the domain at times $0$, $T/4$,
  $T/2$ and $3T/4$, where $T=2\pi/\omega_0$ is the wave period. The
  instant labelled $t=0$ already corresponds here to about 45 wave
  periods, at a time when the beam centred on the attractor has
  reached its final structure.}
\label{fig2}
\end{figure*}

\section{The linear regime}

\subsection{General properties of the steady-state response and comparison with theory}
 \label{sect_linear}

 An illustrative example is shown in Fig.~\ref{fig2} of a direct
 numerical simulation of an inertial wave attractor quickly developing
 in our computational domain. The forcing amplitude is chosen here to
 be small enough ($F_0=7.5\times10^{-4}$) so that the nonlinear terms
 do not play a significant role.
 During a relatively short transient phase
 (a few tens of wave periods), the structure of the wave attractor
 starts to appear, forming a square shape inside our domain. The
 figure shows the oscillatory flow that is established after this
 transient phase. The inertial wave beam, an oscillating shear layer
 formed around the attracting structure, is clearly visible,
 the excitation being optimal in this case. We also see in this
 figure the transverse character of the inertial waves, with the phase
 propagating perpendicular to the beam.


\begin{figure}
\centering
\includegraphics[width=14cm]{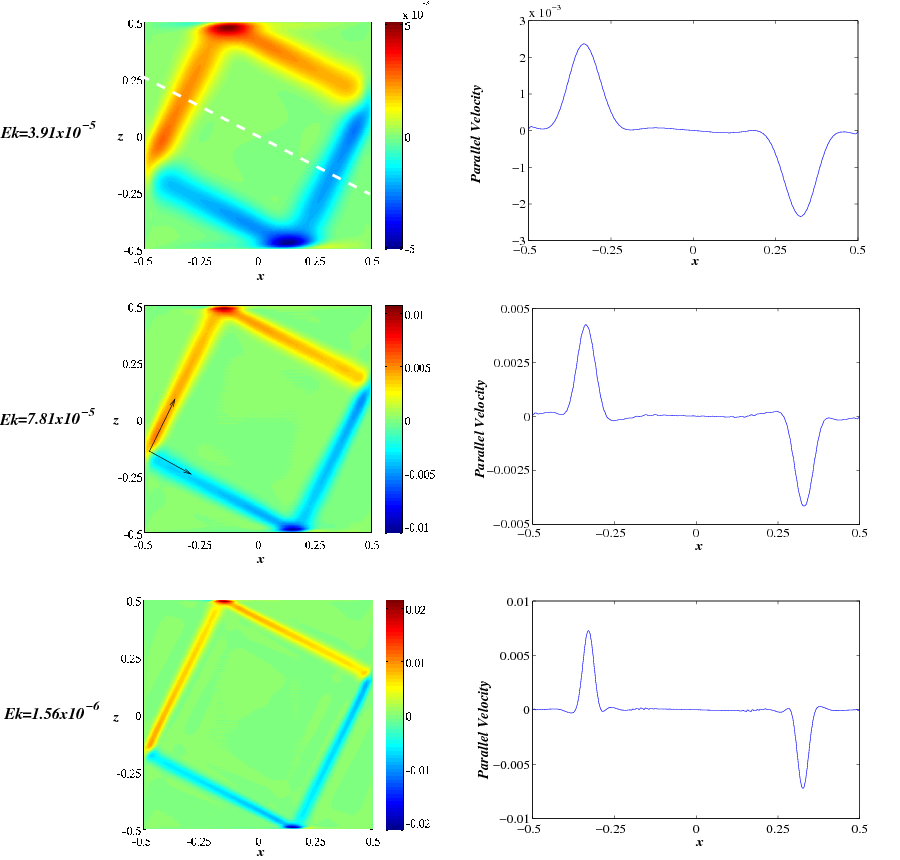}
\caption{Snapshot of the velocity component parallel to the attractor
  after the structure has settled (left panels), and a cut of the same
  velocity (right panels) through the attractor, along the white
  dashed line represented in the top left panel, for 3 different
  values of the Ekman number.}
\label{fig3}
\end{figure}

The first analysis we make is to test in a simple way the predicted
scaling of the thickness of the beam with the viscosity
\citep{Ogilvie05}, namely that the width scales with $Ek^{1/3}$ where
$Ek$ is the Ekman number. We thus study simulations with various Ekman
numbers at a fixed forcing amplitude $F_0=7.5\times10^{-4}$ and measure
the width of the beam in each case. Illustrative results are shown in
Fig.~\ref{fig3}, where we present for three different values of the
Ekman number, a snapshot of the component of the velocity parallel
to the beam. The spatial structure of the attractor then clearly
appears. We add to these snapshots a cut through the attractor of this
parallel velocity for each case. The thickness of the beam is measured
in those various cases (we spanned $Ek$ values from $7.81 \times
  10^{-5}$ to $1.56 \times 10^{-6}$) and we find a good agreement with
  the theoretical scaling of $Ek^{1/3}$.

We now proceed to a detailed comparison of the simulations with the
theoretical predictions, looking at the velocity profiles along
various cuts through the attractor.  We examine both the component
parallel to the attractor, $v_\|$, and the component perpendicular to
the $xz$ plane, $v_y$. The velocity component perpendicular to the
attractor and within the $xz$ plane is much smaller.

In each segment of the attractor, the solution takes the form of an
internal wave beam that spreads self-similarly along its length
because of viscosity \citep[cf.][]{Moore69}.  The scaling properties
of the beam are easily understood if we consider a narrow beam emitted
from a theoretical point source.  Since the group velocity of an
internal wave is proportional to its wavelength and directed
perpendicular to the wavevector, the wave propagates along the beam at
a speed proportional to the width of the beam.  As the width increases
as $t^{1/2}$ because of viscous diffusion, so the distance parallel to
the beam increases as $t^{3/2}$.  Therefore the width and the group
velocity increase with the distance to the power $1/3$.

\begin{figure}
\centering
\includegraphics[width=10cm]{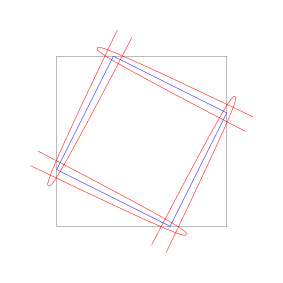}
\caption{Representation of the boundary of the fluid domain (black), the wave
  attractor (blue) and the pattern of self-similarly spreading
  inertial-wave beams in a steady state (red), with virtual sources
  outside the fluid domain.  The width of the beam doubles by viscous
  diffusion along each segment, but is halved at each reflection.}
\label{fig4}
\end{figure}

In a steady state, the viscous spreading of the beam is balanced
globally by the focusing of the beam at its reflection points.  In our
case the width of the beam is halved on each reflection.  It follows
that each segment of the beam has a virtual source that is located
outside the fluid domain at a parallel distance that is $1/7$ of the
length of one side of the attractor.  This ensures that the width of
the beam is doubled between each reflection.  The pattern of beams and
virtual sources is indicated in Fig.~\ref{fig4}.

\citet{Ogilvie05} makes detailed predictions for the streamfunction of
the flow within the plane in the steady-state response in the
asymptotic limit of small Ekman number, where the beam is narrow.  The
`inner solution' found in the neighbourhood of the attractor contains,
in principle, an infinite series of modes with different structures
parallel and perpendicular to the attractor.  Much of the complexity
of the analysis comes from having to determine the amplitudes of the
higher modes, $n\ge1$, which have an oscillatory structure in the direction parallel to the beam.
 The amplitude $a_0$ of the `fundamental' $n=0$
mode is much more easily determined.  Furthermore, it is found that
the $n=0$ mode dominates the response in the case we consider here,
where the body force is very smooth and its curl is independent of
position.  We therefore recall only this component of the asymptotic
inner solution and compare it with the results of our simulations.

Translating the results of \citet{Ogilvie05} into the context and
notations of the present paper, we expect that
\begin{equation}
v_\| = \mathrm{Re} [ U(\tau,\mu) \exp(-i \omega_0 t)],
\label{eq_vpar}
\end{equation}
\begin{equation}
v_y= \mathrm{Re} [ i U(\tau,\mu) \exp(-i \omega_0 t)],
\label{eq_vperp}
\end{equation}
where $\omega_0$ is the forcing frequency (which is also the frequency
of the primary wave), and $U$ is a complex velocity profile given by
\begin{equation}
U(\tau,\mu) = \left(\frac{\omega_0}{\nu \mu} \right)^{1/3} a_0 \,\, \chi_0^{\prime}(\tau).
\end{equation}
Here $\mu$ is the distance along the beam from the virtual source and $\tau$ the transverse distance $\tilde{\tau}$ scaled by $\mu^{-1/3}$ and multiplied by $(\omega_0/\nu)^{1/3}$ to keep the dimension of a distance. The velocity components are scaled with  $\mu^{-1/3}$ to take into account the increase of the group velocity during propagation along the wave beam.
If we consider the leftmost corner of the attractor, the virtual source being located at $x_\mathrm{s},z_\mathrm{s}$, the expressions for $\mu$ and $\tau$ are
\begin{equation}
\mu= (x-x_\mathrm{s}) \sin\beta + (z-z_\mathrm{s}) \cos\beta,
\end{equation}
\begin{equation}
\tilde{\tau}=(x-x_\mathrm{s}) \cos\beta - (z-z_\mathrm{s}) \sin\beta,
\end{equation}
\begin{equation}
\tau=\left(\frac{\omega_0}{\nu \mu} \right)^{1/3} \tilde{\tau},
\end{equation}
where $\beta$ is the angle between the vertical boundary and the beam, shown in Fig.~\ref{fig1}.

The amplitude coefficient of the `fundamental' $n=0$ mode as described above is
\begin{equation}
a_0=\frac{\delta}{\ln \alpha},
\label{eq_a0}
\end{equation}
where $\delta$ and $\alpha$ are respectively the forcing and focusing
constants. The focusing constant $\alpha$ represents the focusing
experienced by a wave beam in one full circuit around the
container. In our case, each reflection at the boundary focuses our
beam by a factor $2$. In one full loop, the beam undergoes four
reflections and the focusing factor is thus $\alpha=2^4=16$.

The factor $\delta$ is related to the change in the streamfunction around each circuit of the wave attractor, due to the body force. In our problem it is
defined in the following way:
\begin{equation}
\delta = \frac{i}{2\omega_0} \int (\boldsymbol{\nabla} \times \boldsymbol{a})_y \,\,\mathrm{d}A,
\end{equation}
where the integral is computed over the area enclosed by the
attractor, and a time-dependence of $\exp(-i\omega_0t)$ is understood
for the forcing.  [This expression differs by a factor of $-2\omega_0$
from the one in \citet{Ogilvie05} because the problem was defined
slightly differently in that paper.]  In our case, it is thus equal to
\begin{equation}
\delta = \frac{i}{2\omega_0} \,\,F_0 \,\,\frac{5 L^2}{9}.
\end{equation}

\begin{figure}
\centering
\includegraphics[width=8cm]{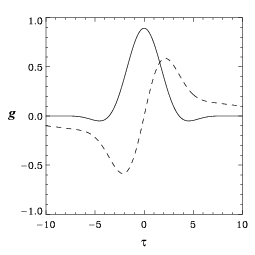}
\caption{Real and imaginary parts (solid and dashed lines) of the
  dimensionless velocity profile of the beam.}
\label{fig5}
\end{figure}

Finally $\chi_0(\tau)$ is the transverse profile of the
streamfunction, which satisfies an ordinary differential
equation whose solutions are obtained by the Laplace transform of
\cite{Moore69}. The first derivative $\chi_0^{\prime}$ which appears
in the expression for the parallel velocity (and also for the
$y$-velocity, which is related to it through the Coriolis force) has
the following integral representation:
\begin{equation}
\chi_0^{\prime}(\tau) = i \int_0^\infty \exp(-i p \tau - p^3)\,\mathrm{d}p = i g(-\tau).
\label{eq_tau}
\end{equation}

Note that the complex function $g(\tau)$ is composed purely
of positive wavenumbers; its real and imaginary parts are plotted in
Figure~\ref{fig5}, and its maximum modulus is
$\max|g|=g(0)=\Gamma({\textstyle\f{4}{3}})\approx0.8930$.

\begin{figure}
\centering
\includegraphics[width=14cm]{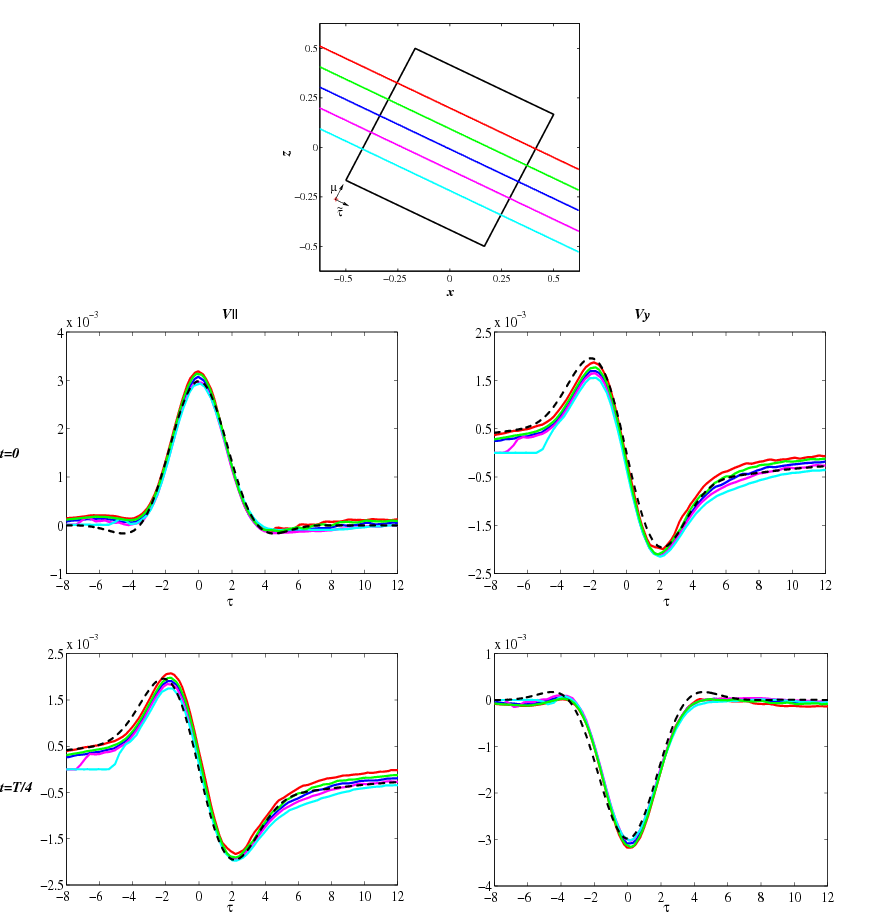}
\caption{Top panel: sketch of the attractor inside the container with a representation of the various cuts made through the attractor, for which the profiles of the parallel and the $y$-velocities are shown.
The red dot at the bottom left of the sketch marks the position of the virtual source, i.e. the origin of the coordinate system $(\tilde{\tau},\mu)$. Four other panels: cuts of the velocities parallel and perpendicular to the beam, as a function of the scaled transverse distance $\tau$ (shown on the top panel). The point $\tau=0$ indicates the centre of the wave beam. The cuts are represented at the same times as the two upper panels of Fig.~\ref{fig2}. The colored lines show the results of the simulations and the dotted black line the theoretical prediction.}
\label{fig6}
\end{figure}

Figure~\ref{fig6} shows cuts at different distances from the virtual source along the beam of the parallel and perpendicular velocities, rescaled by $\mu^{1/3}$ to account for viscous spreading. The cuts are shown as functions of the scaled transverse distance $\tau$, $\tau=0$ corresponding to the centre of the beam. The case shown here is at $Ek=7.81\times10^{-6}$. The theoretical velocities given by expressions (\ref{eq_vpar}) and (\ref{eq_vperp}) and also rescaled by $\mu^{1/3}$ are superimposed on those curves. This figure shows a fairly good agreement between the theoretical profile and the simulated attractor at various temporal phases in the steady-state. We indeed find that both the amplitude of the velocities and the width of the beam scale with the distance to the virtual source to the power $1/3$. In agreement with what was shown in Fig.~\ref{fig2}, the parallel velocity is positive and maximum at temporal phase $t=0$ and changes sign at $\tau=0$ at temporal phase $t=T/4$. It is the opposite for the $y$-velocity, as also expected from the theory. We note that the theoretical profiles are symmetric or antisymmetric with respect to the centre of the beam at these temporal phases. This is verified in the simulations but this is more easily seen for cuts made far from the virtual source (green and red curves) where the left boundary of the domain does not affect the velocity profiles.

\subsection{Building an inertial wave attractor}

\begin{figure}
\centering
\includegraphics[width=14cm]{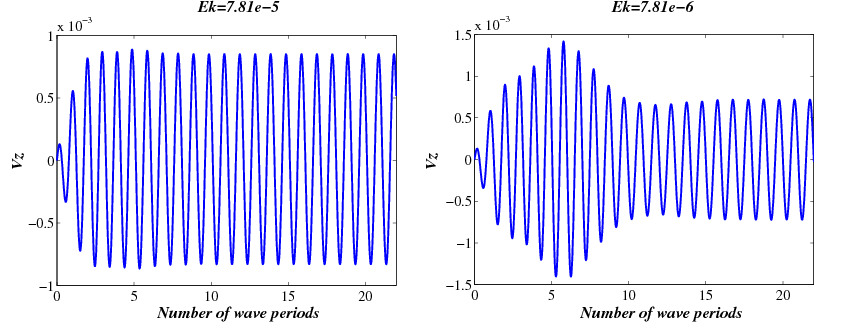}
\caption{Temporal evolution of the vertical velocity $v_z$ at a point located inside the attractor, from the beginning of the simulation, in cases where $Ek=7.81 \times 10^{-5}$ (left panel) and $Ek=7.81 \times 10^{-6}$ (right panel).}
\label{fig7}
\end{figure}

An interesting aspect of the problem that is revealed through direct
numerical simulations of the wave attractor is the transient evolution
from the initial state with zero velocity to the formation of the
final steady-state oscillatory solution described in \cite{Ogilvie05}.
Fig.~\ref{fig7} shows the temporal evolution of the $z$-component of
the velocity inside the attractor from the beginning of the
simulation, for two different values of the Ekman number, namely
$Ek=7.81 \times 10^{-5}$ and $Ek=7.81 \times 10^{-6}$. The transient
evolution is different in these two cases. Since the waves are focused
into a spatial structure of smaller transverse extent in the case
where the viscosity is smaller, and the group velocity of the waves
diminishes in this process, it takes longer to establish the steady
state. To be more precise, in the case where $Ek=7.81 \times 10^{-5}$,
the steady state is reached in about 5 wave periods, while, in the
$Ek=7.81 \times 10^{-6}$ case, we need to wait about twice as long
before the simulation settles. This is consistent with the
scaling of the thickness of the attractor with $Ek$. Since the width
of the beam scales with $Ek^{1/3}$, and the group velocity is
proportional to the wavelength, the time it takes to reach the steady
state also scales with $Ek^{1/3}$. Consequently, if a steady state is
reached in about 5 wave periods in the $Ek=7.81 \times 10^{-5}$ case,
it will be reached in about $5\times 10^{1/3}\approx10.5$ wave-periods in the
$Ek=7.81 \times 10^{-6}$ case, which is consistent with our numerical
results.

If we now focus on the energy balance in the system, we can compare
the energy input by the forcing term in the Navier--Stokes equation
and the energy dissipation due to the viscous term. For the total
kinetic energy to be conserved, these two terms should exactly balance
each other in the steady state, when averaged over one oscillation
cycle. More precisely, we should have in our case where $\nu$ is
constant:
\begin{equation}
\int_{\cal{D}} \,\,\, \frac{\partial}{\partial t}\left(\frac{v^2}{2}\right) \, \mathrm{d}A = \int_{\cal{D}} \,\,\, \boldsymbol{a} \cdot \boldsymbol{v} \, \mathrm{d}A + \nu \int_{\cal{D}} \,\,\, \boldsymbol{v} \cdot \nabla^2 \boldsymbol{v} \, \mathrm{d}A,
\end{equation}
where $\cal{D}$ designates the fluid
  domain.  Equivalently,
\begin{equation}
\frac{\mathrm{d}}{\mathrm{d}t}\int_{\cal{D}} \,\,\, \frac{v^2}{2} \, \mathrm{d}A = \int_{\cal{D}} \,\,\, \boldsymbol{a} \cdot \boldsymbol{v} \, \mathrm{d}A - \nu \int_{\cal{D}} \,\,\, |\boldsymbol{\nabla} \times \boldsymbol{v}|^2 \, \mathrm{d}A,
\label{eq_ek}
\end{equation}
which shows that the rate of change of the total kinetic energy is the
difference of the power input by the body force and the viscous
dissipation rate.

We are here interested in the transient evolution and more precisely
in the time it takes for the power input to reach its steady state
compared to the time it takes for the dissipation rate to settle.
Moreover, in the steady state, the asymptotic value (when $Ek$ tends
to $0$) of the dissipation rate derived in \cite{Ogilvie05} can be
compared to our numerical results. In particular, we wish to
investigate the possible independence of the dissipation rate with
viscosity when $Ek$ tends to $0$.

\begin{figure}
\centering
\includegraphics[width=9cm]{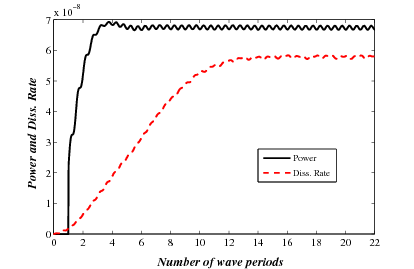}
\caption{Energy input versus dissipation: temporal evolution from the beginning of the simulation, focus on the transient phase.}
\label{fig8}
\end{figure}

We first focus on the simulation at  $Ek=7.81 \times 10^{-6}$ with a forcing amplitude of $F_0=7.5\times10^{-4}$ and compute the 
power input by the force as well as the viscous dissipation rate. Figure~\ref{fig8} shows these quantities. The black line represents the
power input, temporally averaged over a wave period. The red curve represents the evolution of the dissipation rate from the initial state until it reaches a well-established steady state. It is quite clear from this figure that the time-scale for saturation of the power input is much shorter than that of the viscous dissipation. The typical time it takes for the power to reach a steady state is about 4--5 wave periods while the saturation time for the dissipation rate is about 14 periods. The significance of this result is that, while the waves are still in the process of being focused towards the attractor and approaching the scale on which they can be dissipated, the energy flux into the attractor has already reached its asymptotic value. Only later will the energy flux be matched by viscous dissipation in a steady state. 
At that time, a final value of the dissipation rate will be reached, which, for a fixed value of the forcing amplitude $F_0$, should be asymptotically independent of the value of the Ekman number, as shown in \cite{Ogilvie05}.

We note that in our simulation, a discrepancy exists in the saturated regime between the mean power exerted by the force and the dissipation rate. As stated in the first section, this is due to the way the rigid walls are applied in the SNOOPY code, which uses decomposition on Fourier modes and thus usually deals with periodic boundary conditions. The application of a mask on the velocity adds an error of the order of $\Delta t$ in the numerical scheme and thus causes the system to lose part of its energy at the boundaries. Decreasing the time step and increasing the resolution to solve for the viscous boundary layers appearing close to the walls both help to reduce this loss of energy. Various convergence tests were run to have a clear control on the effects of the mask and it was found that the typical time it takes to reach a steady state is not strongly affected; only the final value of the dissipation rate is sensitive to the spatial resolution and time step. In the best cases we have, the energy is conserved at the level of a few percent.

\begin{figure}
\centering
\includegraphics[width=9cm]{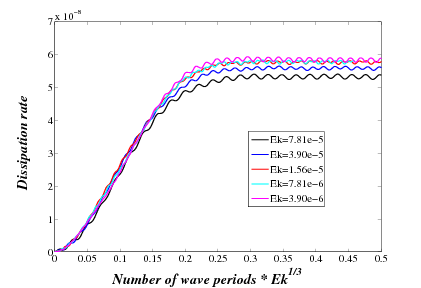}
\caption{Evolution of the average dissipation rate in the domain for various values of the viscosity, as a function of time scaled by $Ek^{1/3}$.}
\label{fig9}
\end{figure}

The value of the viscosity was then varied to investigate the transient phase for different Ekman numbers. The results are shown in Fig.~\ref{fig9}, where only the dissipation rate is represented. It is clear from this plot that the time it takes for the dissipation rate to saturate depends on the viscosity. As stated before, we find that this typical time scales with $Ek^{-1/3}$, a similar scaling to what is found for the width of the final wave beam. However, the long-term dissipation rate seems to approach a limiting value, asymptotically independent of the Ekman number for $Ek\ll1$, as shown by \cite{Ogilvie05}. To have a better comparison to his analytical results, we compute the theoretical value of the asymptotic dissipation rate in the present system. For a smooth forcing function (which we have here), it is shown that the dominant contribution to the dissipation rate is made by the coefficient $a_0$ whose expression was given above (see Eq.~\ref{eq_a0}).
The value of the asymptotic dissipation rate is then obtained by the expression derived by \cite{Ogilvie05}:
\begin{equation}
D \approx 2 \omega_0\, \frac{\pi}{2} \,\,\ln \alpha \,\,\vert a_0 \vert^2 =  \omega_0 \frac{\pi \vert \delta \vert^2}{\ln \alpha}.
\label{eq_diss}
\end{equation}
The factor $2\omega_0$ appears because of the conversion of
notation between the present problem and the one studied by \cite{Ogilvie05}.
In our case here where $F_0=7.5\times10^{-4}$, we find an expected asymptotic value of $6.955 \times 10^{-8}$. In figure~\ref{fig9}, the dissipation rate seems to saturate at a lower value but this is again due to the fact that a small discrepancy exists between the work done by the force and the dissipation rate. If energy conservation was perfect in our domain, the dissipation rate would saturate at the same value as the rate of working. Since the latter is less affected by the application of the mask, we computed the asymptotic value reached by the power input and we obtained $D=6.65 \times 10^{-8}$ for $Ek=7.81 \times 10^{-6}$ and $D=6.74 \times 10^{-8}$ for $Ek=3.9 \times10^{-6}$, thus very close to the asymptotic value found analytically.

\section{Nonlinear cases}

We now investigate the effect of increasing the amplitude of the body force. The idea is that we will reach a level where the nonlinear term will start to play a role and affect the spatial and temporal structure of the attractor.


\subsection{Linear analysis of the instability of an inertial wave beam}

In order to anticipate the behaviour of the nonlinear cases and in particular the characteristics of the possible instabilities which could occur, we carry out a local analysis of a narrow inertial wave beam, such as occurs in the presence of a wave attractor. We consider a simpler problem here, working in an $xy$ plane such that the primary motion is a narrow inertial wave beam aligned with the $x$-axis (i.e. a different coordinate system from the global attractor but without loss of generality or consequences on the characteristics of the instability). The equations considered are the same as Eqs.~\ref{eq_NS} and \ref{eq_div}. We examine the dynamics on
scales comparable to the width of the beam, and therefore assume
periodic boundary conditions in $x$, neglecting the viscous spreading
of the beam along its length.  Consistent with this approximation, we
also neglect the smaller $y$~component of the velocity.  The beam can
then be described as
\begin{equation}
  {\bf \bu}={\bf \bu}_\rmb=u(y,t)\,{\bf \be}_x+w(y,t)\,{\bf \be}_z=\re\{[U(y)\,{\bf \be}_x+W(y)\,{\bf \be}_z]\,\rme^{-\rmi\omega_\rmb t}\},
\end{equation}
\begin{equation}
  p=\re[P(y)\,\rme^{-\rmi\omega_\rmb t}],
\end{equation}
where $U(y)$, $W(y)$ and $P(y)$ are complex amplitudes and
$\omega_\rmb$ is the beam frequency.
The basic equations are satisfied if
\begin{equation}
  -\rmi\omega_\rmb U+2\Omega_yW=0,
\end{equation}
\begin{equation}
  -2\Omega_xW=-\p_yP,
\end{equation}
\begin{equation}
  -\rmi\omega_\rmb W-2\Omega_yU=0,
\end{equation}
which implies $\omega_\rmb=\pm2\Omega_y$, with $W=\pm\rmi U$
respectively.  Nonlinear effects are absent because
${\bf \bu}_\rmb\cdot\bnabla{\bf \bu}_\rmb=\mathbf{0}$.  To sustain the beam
against viscous dissipation, a body force $a=-\nu\nabla^2{\bf \bu}_\rmb$
must be applied.  (In the context of the wave attractor, the viscous
force leads instead to a gradual spreading of the beam along its
length, which is compensated by focusing at the points of reflection.
In the local model, we could allow the beam to decay freely, but
applying a body force has the advantage of maintaining a constant
amplitude in the absence of instability.)

To understand the choice of signs, let us consider the dispersion
relation for 2D inviscid inertial waves,
\begin{equation}
  \omega^2=\f{4(\Omega_xk_x+\Omega_yk_y)^2}{k_x^2+k_y^2},
\end{equation}
which implies a group velocity ${\bf \bv}_\rmg$ given by
\begin{equation}
  2\omega v_{\rmg x}=\f{8\Omega_x(\Omega_xk_x+\Omega_yk_y)}{k_x^2+k_y^2}-\f{8k_x(\Omega_xk_x+\Omega_yk_y)^2}{(k_x^2+k_y^2)^2},
\end{equation}
and similarly for $v_{\rmg y}$.  In the limit $k_x\to0$ relevant for
the beam, we have
\begin{equation}
  v_{\rmg x}\to\f{4\Omega_x\Omega_y}{\omega k_y}.
\end{equation}
If we set up the problem such that $\Omega_x>0$ and $\Omega_y>0$, and
choose the solution with the positive signs, then $\omega_\rmb>0$.  If
the beam is composed of positive wavenumbers $k_y>0$, then $v_{\rmg
  x}$ will be positive and the beam will propagate in the
$+x$~direction.

As stated in Section~\ref{sect_linear}, we assume that the relevant velocity profile of the beam is given by
the dominant $n=0$ mode \citep{Ogilvie05,Moore69,Thomas72} so
that
\begin{equation}
  U(y)=U_0g(\tau)=U_0\int_0^\infty\exp(-k^3+\rmi k\tau)\,\rmd k,
\end{equation}
where $U_0>0$ is the beam amplitude and $\tau=y/L_\rmb$ is a
dimensionless transverse $y$~coordinate, scaled by a characteristic beam width
$L_\rmb$.

In terms of the streamfunction $\psi$ and the relative vorticity
$\omega_z$ of the flow in the $xy$ plane, we have
\begin{equation}
  u_x=\p_y\psi,\qquad
  u_y=-\p_x\psi,\qquad
  \omega_z=\p_xu_y-\p_yu_x=-\nabla^2\psi,
\end{equation}
and the basic equations become
\begin{equation}
  (\p_t+{\bf \bu}\cdot\bnabla)\omega_z-2{\bf \bOmega}\cdot\bnabla u_z=\nu\nabla^2\omega_z+\p_xa_y-\p_ya_x,
\label{basic_omega_z}
\end{equation}
\begin{equation}
  (\p_t+{\bf \bu}\cdot\bnabla)u_z-2{\bf \bOmega}\cdot\bnabla\psi=\nu\nabla^2u_z+a_z.
\label{basic_u_z}
\end{equation}

We linearize these equations about the beam solution, assuming
perturbations $\propto\rme^{\rmi k_xx}$, to obtain
\begin{equation}
  (\p_t+\rmi k_xu)\omega_z'+\rmi k_x\psi'\p_y^2u-2(\Omega_x\rmi k_x+\Omega_y\p_y)u_z'=\nu(-k_x^2+\p_y^2)\omega_z',
\end{equation}
\begin{equation}
  (\p_t+\rmi k_xu)u_z'-\rmi k_x\psi'\p_yw-2(\Omega_x\rmi k_x+\Omega_y\p_y)\psi'=\nu(-k_x^2+\p_y^2)u_z',
\end{equation}
with
\begin{equation}
  \omega_z'=(k_x^2-\p_y^2)\psi',
\end{equation}
where the primes denote the perturbed quantities.
We treat the $y$ direction numerically by taking a Fourier transform
and discretizing the wavenumber $k_y$ on a regular grid, which leads to a large system 
of coupled linear ordinary differential equations (ODEs) in $t$. In practice
this is equivalent to solving the problem in a finite domain of width
$L_y$ with periodic boundary conditions.  Since the beam is then not
isolated but is effectively surrounded by periodic copies, we aim to
make $L_y\gg L_\rmb$ and to seek localized instabilities that do not
rely on the periodic repetition of the beam.

Since $u$ and $w$ are periodic in $t$, this is a Floquet problem. We
integrate the system of ODEs forwards in time over one beam period
$T_\rmb=2\pi/\omega_\rmb$ starting from many different initial
conditions that span the space of initial data, i.e. by making as many computations as the number of ODEs.  
This generates the
monodromy matrix that determines the evolution of an arbitrary initial condition over one beam period.  
We calculate the eigenvalues $\lambda$ of the
monodromy matrix, which are the characteristic multipliers of the
Floquet analysis, and we deduce the complex growth rates $s$ from the
relation $\lambda=\exp(sT_\rmb)$.  The beam is unstable if
$\mathrm{Re}(s)>0$ for any mode.  The frequency of the instability is
given by $\mathrm{Im}(s)$, but this is defined only modulo
$\omega_\rmb$.

\begin{figure}
\includegraphics[width=6cm]{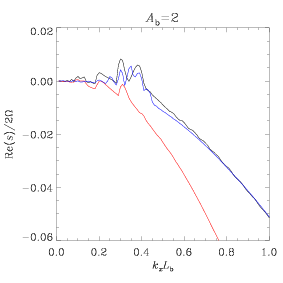}
\includegraphics[width=6cm]{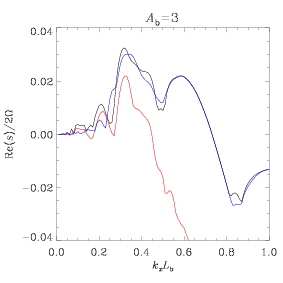}

\includegraphics[width=6cm]{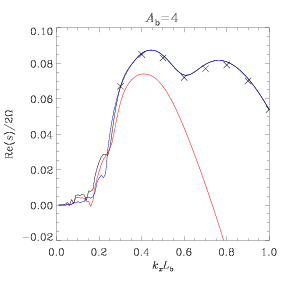}
\includegraphics[width=6cm]{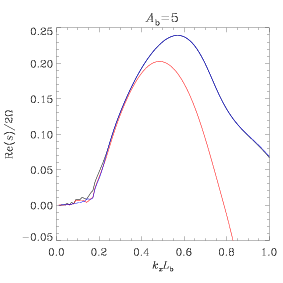}
\caption{Linear stability of beams of different amplitudes.  The
  largest growth rate of the Floquet eigenmodes, in units of
  $2\Omega$, is plotted versus the parallel wavenumber, in units of
  $L_\rmb^{-1}$, for beams of amplitudes $A_\rmb=2$, $3$, $4$ and $5$.
  The black lines are computed for a beam Ekman number of
  $Ek_\rmb=0.05$ and a periodic domain of width $L_y=40\pi\,L_\rmb$.
  The blue lines are for $Ek_\rmb=0.05$ and $L_y/L_\rmb=80\pi$, while
  the red lines are for $Ek_\rmb=0.1$ and $L_y/L_\rmb=40\pi$.  Black
  crosses show growth rates estimated from numerical simulations with
  $Ek_\rmb=0.05$ and $L_y/L_\rmb=40\pi$. Note that the blue and black lines are often indistinguishable, 
indicating convergence of the growth rate with respect to the parameter $L_y/L_\rmb$.}
\label{fig10}
\end{figure}

The problem is determined by several dimensional parameters:
$\Omega_x$, $\Omega_y$, $\nu$, $U_0$, $L_\rmb$, $k_x$ and $L_y$.  We
specialize to the case $\Omega_x=\Omega_y=\Omega/\sqrt{2}$ in which
the beam propagates at an angle $\pi/4$ to the rotation axis, as in
the global problem we consider.  The beam is then described by two
dimensionless parameters, the dimensionless beam amplitude
\begin{equation}
  A_\rmb=\f{U_0}{2\Omega L_\rmb}
\end{equation}
and the beam Ekman number
\begin{equation}
  Ek_\rmb=\f{\nu}{2\Omega L_\rmb^2}.
\end{equation}

The results of the linear calculation can be expressed by finding the
dimensionless growth rate $s/2\Omega$ as a function of the
dimensionless wavenumber $k_xL_\rmb$.  The remaining dimensionless
parameter is $L_y/L_\rmb$, which we aim to make as large as possible.

The maximum relative vorticity in the beam is $0.4514(U_0/L_\rmb)$;
this follows from
$\max|g'|=g'(0)/\rmi={\textstyle\f{1}{3}}\Gamma({\textstyle\f{2}{3}})\approx0.4514$.
Both $x$ and $z$ components of the relative vorticity attain this
maximum value.  The maximum beam vorticity exceeds the vorticity
$2\Omega$ associated with the background rotation when $A_\rmb>2.215$.
(It may also be relevant for stability to compare the components of
the beam vorticity in the direction of the rotation axis with
$2\Omega$; then the beam exceeds the background when $A_\rmb>3.133$, which implies that Rayleigh's criterion is violated near the centre 
of the beam at some phase of the oscillation.)

Sample results of our linear calculations are shown in
Figure~\ref{fig10}.  For sufficiently large $A_\rmb$ the
beam has a robust local instability, for values of $k_xL_\rmb$
somewhat less than unity, that is insensitive to $L_y$ for
$L_y/L_\rmb\gg1$.  The maximum growth rate increases rapidly with
$A_\rmb$.  For smaller $A_\rmb$ weaker instabilities are found in
periodic domains, but these depend significantly on $L_y$ and may
vanish in the limit $L_y/L_\rmb\gg1$.  They probably involve
parametric resonances with global inertial modes of the periodic
domain, and rely on the artificial periodic repetition of the beam.
For $A_\rmb=1$, there is no instability for $Ek_\rmb=0.05$ or $Ek_\rmb=0.1$.

We have verified the linear calculations and determined the nonlinear
outcome of the instability of the beam through direct numerical
simulations in which equations (\ref{basic_omega_z}) and
(\ref{basic_u_z}) are solved in a periodic box using a spectral
method.  The scaled box length $L_x/L_\rmb$ enters as an additional
dimensionless parameter.  The beam is initialized and maintained
through a body force as described above.  A very small random
perturbation is added to the flow, either with a single wavenumber
$k_x$ or a broad spectrum.  The outcome of the simulations is in
excellent agreement with the Floquet analysis regarding the growth
rates of the unstable modes.  Some of these growth rates are indicated
as crosses in Figure~\ref{fig10}.
In cases with a strong instability, the transverse kinetic energy grows and saturates.  The other two parts of the kinetic energy 
are reduced below their laminar values but are still greater than the transverse energy.  
We find that the temporal power spectrum is quite noisy but shows a main peak at the beam frequency and 
two smaller peaks at subharmonics close to half the primary wave frequency.

\subsection{Parameters of the beams expected in the global wave attractor}

In this section we estimate the dimensionless parameters of the
inertial-wave beams expected to occur in the global wave attractor simulations on the
basis of the linear asymptotic analysis of \cite{Ogilvie05}.  If we
neglect the nonlinear terms in the basic equations
(\ref{basic_omega_z}) and (\ref{basic_u_z}), assume a time-dependence
$\propto\rme^{-\rmi\omega_\rmb t}$ and set to zero the force component
$a_z$, then we obtain
\begin{equation}
  -(-\rmi\omega_\rmb-\nu\nabla^2)\nabla^2\psi-2{\bf \bOmega}\cdot\bnabla u_z=F_0,
\end{equation}
\begin{equation}
  (-\rmi\omega_\rmb-\nu\nabla^2)u_z-2{\bf \bOmega}\cdot\bnabla\psi=0,
\end{equation}
where $F_0=\p_xa_y-\p_ya_x$.  We select coordinates aligned with the
attractor, in which one of the beams is centred on the line $y=0$, and
$\omega_\rmb=2\Omega_x=2\Omega_y=\sqrt{2}\Omega$ as in the previous
section.  In this case our equations become
\begin{equation}
  -(-\rmi\omega_\rmb-\nu\nabla^2)\nabla^2\psi-\omega_\rmb(\p_x+\p_y)u_z=F_0,
\end{equation}
\begin{equation}
  (-\rmi\omega_\rmb-\nu\nabla^2)u_z-\omega_\rmb(\p_x+\p_y)\psi=0,
\end{equation}
which may be combined into
\begin{equation}
  (\rmi\omega_\rmb+\nu\nabla^2)^2\nabla^2\psi+\omega_\rmb^2(\p_x+\p_y)^2\psi=(\rmi\omega_\rmb+\nu\nabla^2)F_0.
\end{equation}
The viscous terms are important only in narrow beams.  Expanding out
the equation and retaining only the leading viscous terms, we obtain
\begin{equation}
  2\omega_\rmb^2\p_x\p_y\psi+2\rmi\omega_\rmb\nu\nabla^4\psi=\rmi\omega_\rmb F_0,
\end{equation}
which can be written in the same form as the basic equation~(3.1) of
\cite{Ogilvie05},
\begin{equation}
  \rmi\p_x\p_y\psi+\epsilon^3\nabla^4\psi=f,
\end{equation}
if we identify
\begin{equation}
  \epsilon^3=-\f{\nu}{\omega_\rmb},\qquad
  f=-\f{F_0}{2\omega_\rmb}.
\end{equation}

As stated in Sect.~\ref{sect_linear} and according to \cite{Ogilvie05}, the dominant `$n=0$' part of the asymptotic inner
solution close to the attractor is
\begin{equation}
  \psi\sim a_0\chi_0(\tau)+b_0,
\end{equation}
where $b_0$ is an unimportant constant and $a_0$ and $\chi_0(\tau)$ were defined above (see Eqs~\ref{eq_a0} and \ref{eq_tau}).
We recall that:
\begin{equation}
  a_0=\f{\delta}{\ln\alpha},\qquad
  \delta=-\rmi\int_\mathrm{attractor}f\,\rmd A,
\end{equation}
that the function $\chi_0(\tau)$ is
related to our function $g(\tau)$ through
\begin{equation}
  \chi_0'(-\tau)=\rmi g(\tau),
\end{equation}
and that the scaled coordinate $\tau$ is defined as $\tau=y/L_\rmb$ where

\begin{equation}
  L_\rmb=-\epsilon\mu^{1/3}=\left(\f{\nu \mu}{\omega_\rmb}\right)^{1/3},
\end{equation}
where $\mu$ is the distance along the beam from its virtual source, outside
the fluid domain.

 We therefore expect a velocity parallel to the beam
of amplitude 
\begin{equation}
  U_0=-\f{\rmi a_0}{L_\rmb}=\f{F_0}{2\omega_\rmb L_\rmb\ln\alpha}\f{5L^2}{9}.
\end{equation}

The predicted dimensionless beam amplitude is then
\begin{equation}
  A_\rmb=\f{U_0}{2\Omega L_\rmb}=\f{1}{2^{5/6}4\ln2}\left(\f{5}{9}\right)^{2/3}\left(\f{\nu}{2\Omega L^2}\right)^{-2/3}\f{F_0}{(2\Omega)^2}\left(\f{\mu}{L_\rma}\right)^{-2/3}
\end{equation}
and the beam Ekman number is
\begin{equation}
  Ek_\rmb=\f{\nu}{2\Omega L_\rmb^2}=\left(\f{9}{10}\right)^{1/3}\left(\f{\nu}{2\Omega L^2}\right)^{1/3}\left(\f{\mu}{L_\rma}\right)^{-2/3}.
\end{equation}
$L_\rma$ is the length of one side of the attractor and is thus equal in the complete problem to $\sqrt{5/9} \, L$. Note that the dimensionless distance $\mu/L_\rma$ along the beam ranges from $1/7$ to $8/7$, already showing that if the instability develops, it should start where the dimensionless amplitude is maximal and thus at the reflection points where $\mu/L_\rma=1/7$. This is indeed where the subharmonics in our global attractor simulations seem to be excited first.

\begin{figure}
\centering
\includegraphics[width=12cm]{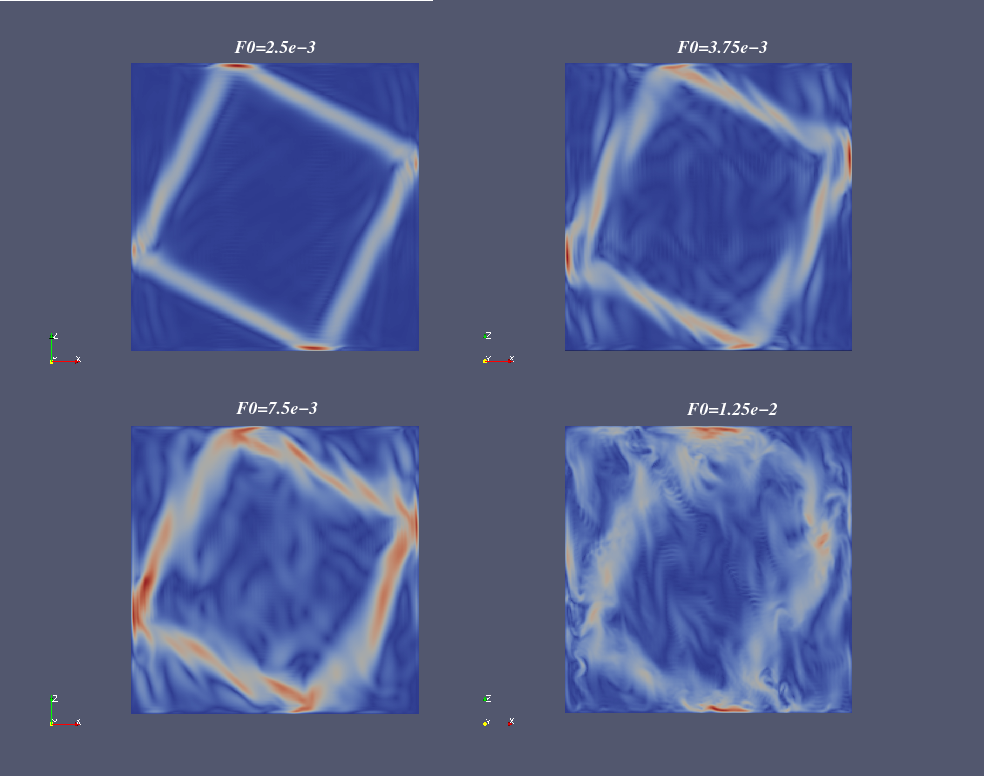}
\caption{Snapshot of the norm of the velocity in the plane $\sqrt{v_x^2+v_z^2}$ in the statistically steady state for four different forcing amplitudes, from $F_0=2.5\times10^{-3}$ to $F_0=1.25\times10^{-2}$.}
\label{fig11}
\end{figure}

\subsection{Instability of the primary wave attractor in the full non-linear simulations}

Let us now come back to the direct numerical simulations of the global attractor which was studied in Section \ref{sect_linear} in the linear regime. We are thus returning to the coordinate system defined in Section \ref{sect_intro}, in which coordinates $(x,z)$ are aligned with the container and $y$ is in the third direction. The two previous subsections indicate that an instability should be clearly visible when the equivalent beam Ekman number is around $Ek_b=0.05$. We choose to fix the value of $Ek={\nu}/{2\Omega L^2}$ to a value which was studied in the linear regime (i.e. $Ek=3.9\times10^{-6}$), which, according to the previous subsection, translates into a beam Ekman number of $Ek_b\approx0.0556$ at the reflection point where $\mu/L_\rma=1/7$.

For this value of $Ek_\rmb$, the linear analysis above shows us that the first signs of a clear local instability should be seen for a beam amplitude of $A_\rmb\approx4$, leading to a forcing amplitude around ${F_0}/{(2\Omega)^2}=2\times10^{-3}$. Indeed, for $A_\rmb=1$, all the growth rates are negative and the threshold for instability is not reached yet. As shown in Fig.~\ref{fig10}, when $A_\rmb=2$, the growth rates of possible unstable modes are extremely small (of the order of $70$ wave periods) and the instability thus likely not to be seen in our global simulations. On the contrary when we reach the value of $A_\rmb\approx4$, a clear instability is present in the linear analysis as seen in the two bottom panels of Fig.~\ref{fig10}. We have not tried to determine precisely the threshold for instability in the global simulations, but we wish to determine whether we get some clear signs of instability when we force the system with ${F_0}/{(2\Omega)^2} \geq 2\times10^{-3}$.

We show in Fig.~\ref{fig11} the norm of
the velocity $(v_x,0,v_z)$ in the $(x,z)$-plane for increasing values
of the forcing amplitude, for an Ekman number
$Ek=3.9\times10^{-6}$. The four snapshots were chosen at a time when
both the power input and the dissipation rate have reached a
statistically steady state and at a temporal phase close to $t=0$
(i.e. when the parallel velocity is positive and maximum). As seen in
this figure, as the forcing amplitude is increased, the spatial
structure of the beam is perturbed and starts to be modified mostly at
the locations of reflections, i.e. at the corners of the attractor. At
$F_0=3.75\times10^{-3}$ and $F_0=7.5\times10^{-3}$, we start to
distinguish waves propagating at different angles with respect to the
rotation vector, indicating the potential existence of excitation at
different frequencies, as will be presented later. In the case with
the largest forcing amplitude ($F_0=1.25\times10^{-2}$), the attractor
is hardly visible since the kinetic energy is much less concentrated
than in the previous cases. The wave focusing is in this case less
efficient. The primary wave beam is nonetheless still distinguishable
but tends to be broader and more affected by the surrounding velocity
field.  It is then clear that the attractor is strongly affected by
the nonlinearities and filters will have to be applied to distinguish
the spatial structure of the beam in these cases. We make use of the
procedure of \cite{Grisouard08} to correctly visualize the beam in the
nonlinear cases. We choose to filter the $z$ and the $y$-components of
the velocity at the frequency of the forcing term first and then at
various subharmonics to observe the spatial structure at each of those
frequencies. Our filtered field at frequency $\omega$ thus takes the
following expression:

\begin{equation}
{\bf V}_{\omega}(x,z)=\frac{\omega}{N\pi} \,\, \int_{t_\mathrm{i}}^{t_\mathrm{f}} \,\, {\bf v}(x,z,t) \exp[\mathrm{i}\, \omega (t-t_\mathrm{i})]\, \mathrm{d}t.
\end{equation}
where $t_\mathrm{i}$ is taken in the well-established steady state at
about $500$ wave periods $(T=2\pi/\omega_0)$ after initialization and
$t_\mathrm{f}$ is taken $N\approx100$ wave periods later. 

\begin{figure}
\centering
\includegraphics[width=14cm]{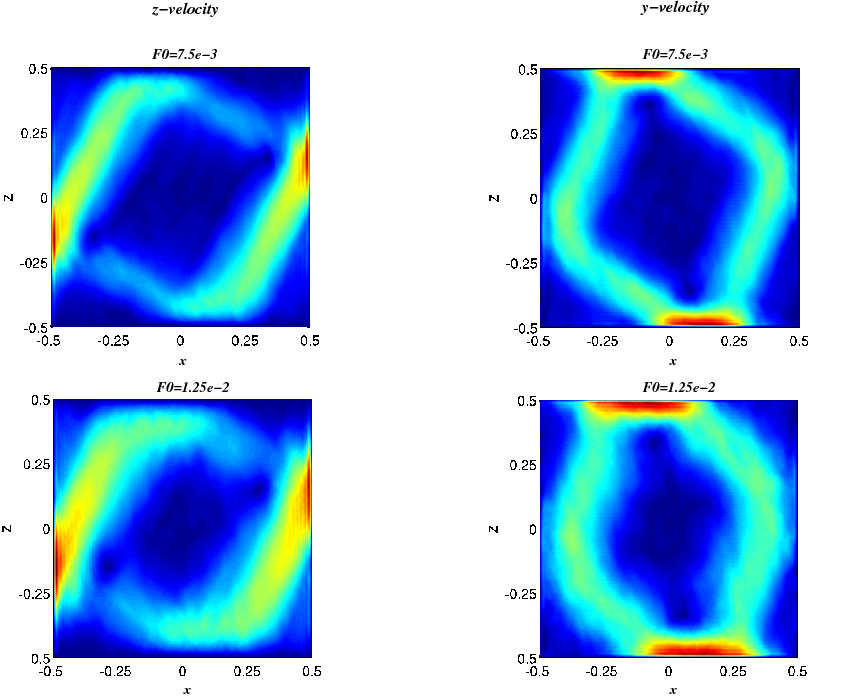}
\caption{Modulus of the filtered velocity field at $\omega_0=\sqrt{2}\Omega$ for the cases where the amplitude of the forcing is equal to $F_0=7.5\times10^{-3}$ and $F_0=1.25\times 10^{-2}$ and with $Ek=3.9\times10^{-6}$.}
\label{fig12}
\end{figure}

Figure~\ref{fig12} shows the result of the filtering of $v_z$ and $v_y$
at the forcing frequency $\omega_0$ in two cases, namely when the
forcing amplitude is set to $7.5\times10^{-3}$ and when it is set to
$1.25\times10^{-2}$. The latter case corresponds to the last panel of
Fig.~\ref{fig11} where the beam was clearly strongly perturbed by the
nonlinearities. The filtering procedure thus enables us to recover the
spatial structure of the primary beam, \emph{i.e.} at the forcing
frequency. One striking feature here is that the effective thickness
of the attractor increases in the nonlinear cases, indicating that
instability and nonlinear dynamical processes prevent the waves from
focusing on very short scales.

When the energy balance is
investigated, we find again in those nonlinear cases that in the
statistically steady state, the dissipation rate is on average equal
to the power input from the body force, with an error of a few percent
that may be due to the effects of the mask as in the linear regime.

\begin{figure}
\centering
\includegraphics[width=10cm]{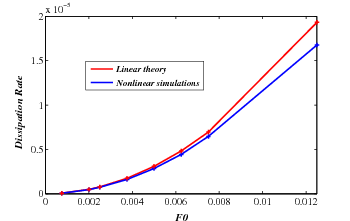}
\caption{Blue curve: dissipation rate computed for various forcing amplitudes (from \ref{eq_ek}). Red curve: dissipation rate predicted by linear theory (from \ref{eq_diss}), which scales with $F_0^2$.}
\label{fig13}
\end{figure}

However, there is a difference in the nonlinear cases in the time it
takes for the dissipation rate to saturate to its final value.
Indeed, when the forcing amplitude is increased, it takes less time
for the dissipation rate to saturate, which is related to the fact
that the final beam becomes broader, as seen in Fig.~\ref{fig12}.  When
the forcing is stronger, it thus seems that the kinetic energy reaches
the dissipative scales more rapidly by an instability mechanism than
it does purely by geometrical focusing. As a consequence, the level of
saturation for the dissipation rate is reached before the waves have
time to focus on the attractor and the spatial
structure of the wave beam is strongly modified.  The dissipation rate
is computed for the cases with $F_0=7.5\times10^{-4}, 2\times10^{-3},
10^{-2}, 3.75\times10^{-3}, 5\times10^{-3}, 6.25\times10^{-3},
7.5\times10^{-3}$ and $1.25\times10^{-2}$ and compared to the
asymptotic linear value calculated in the previous section.
Figure~\ref{fig13} shows the values of the dissipation rate as a
function of the forcing amplitude $F_0$ (blue curve) and the
asymptotic values found in the linear theory (red curve). We find an
agreement between the two curves of around $95\%$ for the first three
points, where the nonlinear effects are limited. The agreement then
decreases to $92\%$ for the four following values and then drops to
$85\%$ for the last case which corresponds to
$F_0=1.25\times10^{-2}$. In the latter case, the strong nonlinear
effects prevent the waves from focusing on the short scale of the
theoretical attractor. The instability of the wave beam provides here
another way to transfer energy to smaller and smaller wavelengths
until they reach the dissipative scale.

\subsection{Non linear outcome of the instability in the global simulations}

\begin{figure}
\centering
\includegraphics[width=10cm]{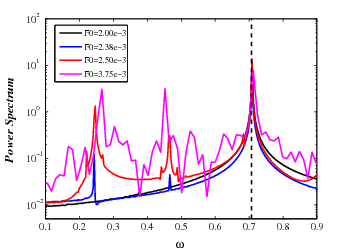}
\caption{Power spectrum of the kinetic energy for cases at various forcing amplitudes, as a function of frequency $\omega$ and for $Ek=3.9\times10^{-6}$. The vertical dashed line shows the location of the primary wave frequency $\omega_0=\sqrt{2}\Omega=\sqrt{2}/2$.}
\label{fig14}
\end{figure}

Since it is clearly visible in Fig.~\ref{fig11} that the structure of
the flow changes with increasing amplitude of the forcing applied to
the system, we need to focus on the temporal power spectrum of the
velocity, in order to have a precise idea of which frequencies are
excited in the domain. Fig.~\ref{fig14} shows the spectra of one
component of the velocity ($v_z$) at a particular point in the
attractor, for a forcing amplitude varying from $F_0=2\times10^{-3}$
to $F_0=3.75\times10^{-3}$ and keeping $Ek=3.9\times10^{-6}$. All spectra are superimposed so that we can
clearly analyse which frequencies appear for which forcing amplitude.
In the linear case at $F_0=2\times10^{-3}$, the only visible peak is
the one at the forcing frequency $\omega=\omega_0=\sqrt{2}/2$. This peak of highest amplitude is present for all cases we
have computed. This also explains why the spatial structure of the
attractor could still be recovered even for the relatively high forcing
amplitude of $F_0=1.25\times 10^{-2}$.

When the forcing is increased and thus when the simulations become more and more nonlinear, additional frequencies are excited, especially at lower values. 
It seems that the instability here transfers energy from the primary wave to two recipient waves at lower frequencies. This was also found in laboratory experiments where internal gravity waves were excited \citep{Joubaud12}.

As the forcing amplitude increases and the nonlinearities become even more important, these subharmonics tend to get closer to half the primary wave frequency. The resonance condition for the frequencies remains true in those more nonlinear cases. 
This is mostly visible in the case where $F_0=3.75\times10^{-3}$ in which the peaks at low frequencies move closer to $\omega_0/2=\sqrt{2}/4$. This can be explained by the fact that as the forcing amplitude is increased, the Reynolds number of the flow becomes larger and the instability behaves as if the viscosity was reduced. The characteristics of the instability thus become closer and closer to the inviscid case: shorter secondary waves with a frequency close to half the primary wave frequency, in agreement with the dispersion relation of such inertial waves.

\begin{figure}
\centering
\includegraphics[width=14cm]{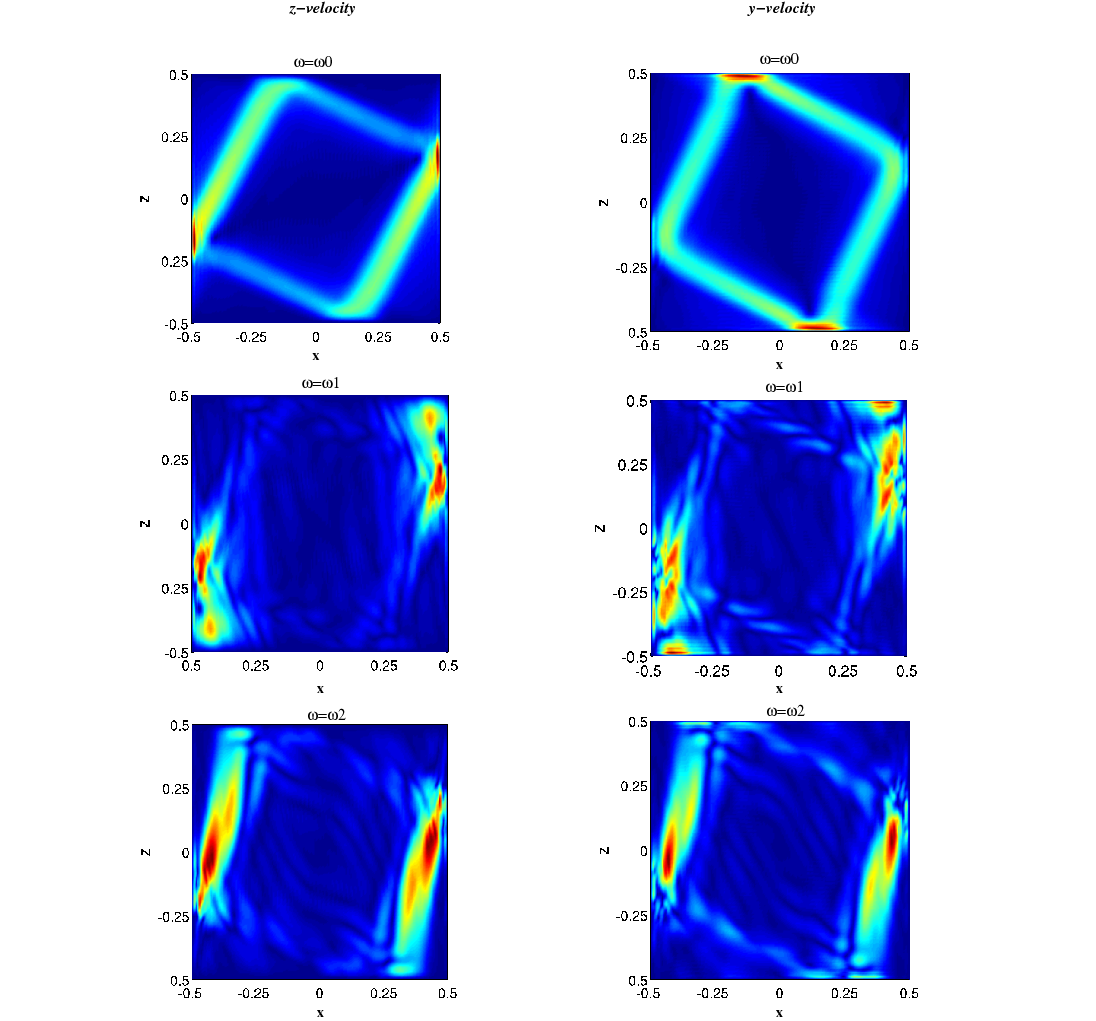}
\caption{Filtered velocity field ($z$ and $y$ components) for the case with a forcing amplitude of $F_0=2.5\times10^{-3}$. The filtering is made at the 3 principal frequencies visible in the power spectrum, namely $\omega_0=\sqrt{2}/2$, $\omega_1=0.244$ and $\omega_2=0.463$.}
\label{fig15}
\end{figure}

It is then interesting to use the same procedure as the previous section to filter the velocity field not at the primary frequency $\omega_0$ like we did to get the spatial structure of the main attractor, but at the subharmonic frequencies, namely at $\omega_1$ and $\omega_2$. The results are shown in Fig.~\ref{fig15} for the case at $F_0=2.5\times10^{-3}$, where the frequencies of the subharmonics clearly appear in the power spectrum of Fig.~\ref{fig14}.

In Fig.~\ref{fig15}, six panels are represented, showing the filtering of the $z$ and the $y$-components of the velocity at the frequency of the primary wave $\omega_0$, and at the frequencies where the Fourier transform shows a signal, namely at $\omega_1\approx0.244$, $\omega_2\approx0.463$. Those values correspond approximately to $\omega_1/\omega_0=0.345$ and $\omega_2/\omega_0=0.655$. On the top panels, the regular spatial structure of the initial attractor is recovered, with a typical thickness slightly larger than in the linear simulations. On the next two panels, it seems that the secondary waves are excited within the primary wave attractor in a region where the energy of the primary wave is maximum (i.e. close to the walls where the reflections occur). The secondary waves then propagate with a different angle with respect to the rotation vector compared to the primary wave. This is due to the dispersion relation of such inertial waves which implies that the angle between the rotation vector and the direction of propagation depends directly on the wave frequency. Since those waves have smaller wavelengths than the primary wave, they also dissipate more quickly and thus do not propagate very far away from their region of excitation. This is especially true for the waves excited at $\omega_1$. Indeed, the mid-panels show a strong location of kinetic energy at this frequency close to the region of excitation but the structure stays spatially localized very close to the corners of the attractor. This is less true for the waves excited at the higher frequency $\omega_2$ which clearly exhibit a region of propagation along a beam inclined at an angle $\theta_2$ with respect to the rotation vector, such that $\sin\theta_2=\omega_2/2\Omega$.

\section{Summary and conclusion}

The purpose of this work was to simulate numerically the evolution in 2D of an inertial wave attractor in a square container in the linear and nonlinear regimes, with a particular focus on the build-up of the attractor before it reaches a steady-state and the instability occurring in the nonlinear case. This study complements previous work on this problem, namely laboratory experiments and linear calculations. In the linear regime, the characteristics of internal wave attractors predicted by theory are well recovered. In particular, we find that the thickness of the attractor scales with the Ekman number to the power $1/3$. For a fixed Ekman number, we moreover confirm that after a reflection at the boundary, the amplitude of the velocity field decreases as the distance from the source to the power $1/3$ while the breadth of the wave beam increases by the same factor. Finally, we find in the simulations that, for a body force of fixed amplitude, the dissipation rate indeed becomes independent of the Ekman number and approaches the asymptotic value derived analytically in \cite{Ogilvie05}.

Since we compute the evolution of the wave attractor from the very beginning where an incompressible fluid is forced in a closed container, we have access to the transient phase before the system reaches a statistically steady state. In this simple linear case, only two quantities will play a role in the energy balance: the power input due to the periodic forcing of the system and the dissipation rate. Both those quantities reach the same absolute value in the steady state in order for energy to be conserved. However, we find here that the power (i.e. the rate of working by the applied force) saturates in about 5 wave periods in all cases, while the dissipation rate reaches its asymptotic value on a timescale that depends on the Ekman number, typically close to 15 wave periods. The kinetic energy of the flow therefore continues to build up, and the dissipation rate increases, until the wave attractor has reached its final thickness, while the rate of working by the applied force saturates much earlier. This is consistent with the viewpoint that the attractor acts as a ``black hole'' that absorbs the energy flux associated with the quasi-inviscid, large-scale waves that are focused towards it.

The present work also focused on the properties of inertial wave attractors in the nonlinear regimes. To do so, the forcing amplitude was increased until nonlinear effects would start to play a significant role. The primary wave beam oscillating at the forcing frequency $\omega_0$ is then found to be subject to an instability which is investigated through linear analysis and nonlinear calculations of the simpler case of a single infinite narrow wave beam. The onset of this instability and the nonlinear outcome are determined in this simpler case and are consistent with the characteristics of the global wave attractor in the nonlinear regime. It is found that this instability transfers energy to waves of lower frequencies and shorter wavelengths. Those secondary inertial waves are excited close to the reflection points and then propagate with a direction in agreement with the typical inertial wave dispersion relation. Since they possess a smaller scale, they dissipate quickly and propagate over much shorter distances than the primary waves. As the forcing amplitude is increased, their effective Reynolds number is increased and the frequencies of the secondary waves drift towards $\omega_0/2$, thus approaching the outcome of a regular parametric subharmonic instability in the inviscid case. In the cases where instability occurs, the energy is transferred rapidly to shorter wavelengths which then dissipate. As a consequence, the dissipation rate reaches its saturation more quickly than in the linear cases. In the present work, this type of instability is described theoretically and recovered in nonlinear direct numerical simulations. An instability of a wave beam with similar characteristics was found in an experiment and supported by nonlinear numerical simulations by \cite{Clark10}. Other types of instabilities were also observed in laboratory experiments of gravity waves \citep{Joubaud12} or inertial waves \citep{Bordes12} excited in closed containers. 

In geophysical or astrophysical situations, inertial waves can be excited in stars or planets by tidal forcing due to an orbiting companion. Moreover, these systems can easily be in a nonlinear regime if the forcing is sufficiently strong (because of the companion being sufficiently close and/or massive). Under these conditions, wave attractors can still be present but the flow may be unstable and the dissipation rate and tidal torque may be reduced below the value predicted by linear theory. 

Of course the problem considered here was 2D, consisting of a very simple geometry and it has been shown that moving to 3D and spherical geometry may strongly modify the behaviour of wave attractors and the characteristics of the dissipation processes \citep{Rieutord10}. However, it is the first example here of strong nonlinear effects emerging in simulations of wave attractors, namely subharmonic instabilities. It is interesting to note that our simulations have counterparts in the recent fluid experiments of \cite{Scolan13}, which also exhibit a strongly nonlinear regime of wave attractors. 

Obviously, much more can be done on this problem. If we consider the conditions in the interiors of stars or giant planets, large-scale flows such as meridional flows or differential rotation are likely to exist and modify the build-up of potential wave attractors. Turbulent convection will of course also probably act on the wave focusing and defocusing and may disrupt the formation of wave attractors. A next possible step could be the computation of the same kind of simulations for a compressible fluid in a rotating spherical shell, where both gravity and inertial waves would exist. In spherical geometry, critical latitude singularities were shown to be dominant features of periodically forced flows in a linear regime. It would be valuable to carry out fully nonlinear simulations of these phenomena. 

\acknowledgements
The authors are very pleased to acknowledge fruitful discussions with Geoffroy Lesur about the SNOOPY code and its use for this particular problem.

\bibliographystyle{jfm}

\bibliography{jouv_ogil_2014_arxiv}

\begin{thebibliography}{14}
\expandafter\ifx\csname natexlab\endcsname\relax\def\natexlab#1{#1}\fi

\bibitem[Bajars et al.(2013)]{Bajars13}
{\sc Bajars, J., Frank, J., Maas, L. R. M.} 2013 On the appearance of internal wave attractors due to an initial or parametrically excited disturbance {\em J.~Fluid Mech.\/}, {\bf 714}, 283--311.

\bibitem[Baruteau \& Rieutord(2013)]{Baruteau13}
{\sc Baruteau C. \& Rieutord, M.} 2013 Inertial waves in a differentially rotating spherical shell {\em J.~Fluid Mech.\/}, {\bf 719}, 47--81.

\bibitem[Bordes et al.(2012)]{Bordes12}
{\sc Bordes, G., Moisy, F., Dauxois, T. \& Cortet P.-P.} 2012 Experimental evidence of a triadic resonance of plane inertial waves in a rotating fluid {\em Physics of fluids \/}, {\bf 24}, 014105.

\bibitem[Clark \& Sutherland(2010)]{Clark10}
{\sc Clark, H. A. \& Sutherland, B. R.} 2010 Generation, propagation, and breaking of an internal wave beam {\em Physics of fluids \/}, {\bf 22}, 076601.

\bibitem[Dintrans et al.(1999)]{Dintrans99}
{\sc Dintrans, B., Rieutord, M. \& Valdettaro, L.} 1999 Gravito-inertial waves in a rotating stratified sphere or spherical shell {\em J.~Fluid Mech.\/}, {\bf 398}, 271--297.

\bibitem[Echeverri et al.(2011)]{Echeverri11}
{\sc Echeverri, P., Yokossi, T., Balmforth, N. J. \& Peacock, T.} 2011 Tidally generated internal-wave attractors between double ridges {\em J.~Fluid Mech.\/}, {\bf 669}, 354--374.

\bibitem[Grisouard et al.(2008)]{Grisouard08}
{\sc Grisouard N., Staquet C. \& Pairaud, I. } 2008 Numerical simulation of a two-dimensional internal wave attractor {\em J.~Fluid Mech.\/}, {\bf 614}, 1--14.

\bibitem[Hazewinkel et al.(2008)]{Hazewinkel08}
{\sc Hazewinkel, J. van Breevoort, P., Dalziel, S. B., Maas, L., R., M.} 2008 Observations on the wavenumber spectrum and evolution of an internal wave attractor {\em J.~Fluid Mech. \/}, {\bf 598}, 373--382.

\bibitem[Hazewinkel et al.(2011a)]{Hazewinkel11a}
{\sc Hazewinkel,J., Grisouard, N. \& Dalziel, S. B.} 2011 Comparison of laboratory and numerically observed scalar fields of an internal wave attractor {\em European Journal of Mechanics \/}, {\bf 30}, 51--56.

\bibitem[Hazewinkel et al.(2011b)]{Hazewinkel11b}
{\sc Hazewinkel,J., Maas, L. R. M. \& Dalziel, S. B.} 2011 Tomographic reconstruction of internal wave patterns in a paraboloid {\em Experiments in fluids \/}, {\bf 50}, 247--258.

\bibitem[Joubaud et al.(2012)]{Joubaud12}
{\sc Joubaud, S., Munroe, J., Odier, P. \& Dauxois, T.} 2012 Experimental parametric subharmonic instability in stratified fluids {\em Physics of Fluids\/} {\bf 24}, 041703.

\bibitem[Lam \& Maas(2008)]{Lam08}
{\sc Lam, F.-P. A. \& Maas L. R. M.} 2008 Internal wave focusing revisited; a reanalysis and new theoretical links {\em  Fluid Dynamics Research \/}, {\bf 40}, 95--122.

\bibitem[Lesur \& Longaretti(2005)]{Lesur05}
{\sc Lesur G. \& Longaretti P.-Y.} 2005 On the relevance of subcritical hydrodynamic turbulence to accretion disk transport {\em Astronomy \& Astrophysics\/}, {\bf 444}, 25--44.

\bibitem[Lesur \& Longaretti(2007)]{Lesur07}
{\sc  Lesur G. \& Longaretti P.-Y.} 2007 Impact of dimensionless numbers on the efficiency of magnetorotational instability induced turbulent transport {\em MNRAS\/}, {\bf 378}, 1471--1480.

\bibitem[Lesur \& Ogilvie(2010)]{Lesur10}
{\sc  Lesur G. \& Ogilvie, G. I.} 2010 On the angular momentum transport due to vertical convection in accretion discs {\em MNRAS \/}, {\bf 404}, 64--68.

\bibitem[Maas \& Lam(1995)]{Maas95}
{\sc Maas L. R. M., Lam, F.-P. A.} 1995 Geometric focusing of internal waves {\em J.~Fluid Mech.\/}, {\bf 300}, 1--41.

\bibitem[Maas et al.(1997)]{Maas97}
{\sc Maas L. R. M., Benielli, D., Sommeria, J. \& Lam, F.-P. A.} 1997 Observation of an internal wave attractor in a confined, stably stratified fluid {\em Nature \/}, {\bf 388}, 557--561.

\bibitem[Maas(2001)]{Maas01}
{\sc Maas L. R. M.} 2001 Wave focusing and ensuing mean flow due to symmetry breaking in rotating fluids {\em J.~Fluid Mech.\/}, {\bf 437}, 13--28.

\bibitem[Maas \& Harlander(2007)]{Maas07}
{\sc Maas L. R. M. \& Harlander, U.} 2007 Equatorial wave attractors and inertial oscillations {\em  J.~Fluid Mech. \/}, {\bf 570}, 47--67.

\bibitem[Manders \& Maas(2003)]{Manders03}
{\sc Manders, A. M. M. \& Maas L. R. M.} 2003 Observation of inertial waves in a rectangular basin with one sloping boundary {\em  J.~Fluid Mech. \/}, {\bf 493}, 59--88.

\bibitem[Moore \& Saffman(1969)]{Moore69}
{\sc  Moore, D. W. \& Saffman, P. G.} 1969 The structure of free vertical shear layers in a rotating fluid and the motion produced by a slowly rising body {\em Phil. Trans. R. Soc. Lond. \/}, {\bf 264}, 597--634.

\bibitem[Ogilvie(2005)]{Ogilvie05}
{\sc Ogilvie G. I.} 2005 Wave attractors and the asymptotic dissipation rate of tidal disturbances {\em J.~Fluid Mech.\/} {\bf 543}, 19--44.

\bibitem[Ogilvie(2009)]{Ogilvie09}
{\sc Ogilvie G. I.} 2009 Tidal dissipation in rotating fluid bodies: a simplified model {\em MNRAS\/} {\bf 396}, 794--806.

\bibitem[Ogilvie \& Lin(2004)]{Ogilvie04}
{\sc Ogilvie, G. I. \& Lin, D. N. C.} 2004 Tidal dissipation in rotating giant planets {\em Astrophys.~J.\/} {\bf 610}, 477--509.

\bibitem[Rieutord \& Valdettaro(1997)]{Rieutord97}
{\sc Rieutord, M. \& Valdettaro, L.} 1997 Inertial waves in a rotating spherical shell {\em J.~Fluid Mech.\/}, {\bf 341}, 77--99.

\bibitem[Rieutord et al.(2001)]{Rieutord01}
{\sc Rieutord, M., Georgeot, B. \& Valdettaro, L.} 2001 Inertial waves in a rotating spherical shell: attractors and asymptotic spectrum {\em J.~Fluid Mech.\/}, {\bf 435}, 103--144.

\bibitem[Rieutord \& Valdettaro(2010)]{Rieutord10}
{\sc Rieutord, M. \& Valdettaro, L.} 2010 Viscous dissipation by tidally forced inertial modes in a rotating spherical shell {\em J.~Fluid Mech.\/}, {\bf 643}, 363--394.

\bibitem[Scolan et al.(2013)]{Scolan13}
{\sc Scolan H., Ermanyuk E. \& Dauxois T.} 2013 Nonlinear Fate of Internal Wave Attractors {\em Phys.~Rev.~Lett.\/}, {\bf 110}, 234501.

\bibitem[Thomas \& Stevenson(1972)]{Thomas72}
{\sc Thomas, N. H. \& Stevenson, T. N.} 1972 A similarity solution for viscous internal waves {\em J.~Fluid Mech.\/}, {\bf 54}, 495--506.

\end{thebibliography}

\end{document}